%% file: main.tex
\documentclass[a4paper,11pt]{article}
\usepackage{pos}

\usepackage[T1]{fontenc}
\usepackage{units}
\usepackage{wrapfig}
\usepackage{tikz}
\usetikzlibrary{calc}
\usepackage{csquotes}
\tikzset{
	font=\footnotesize,
}
\usepackage{bbold}
\usepackage{bm}
\usepackage{pgfplots}
\pgfplotsset{
	compat=1.16,
	/pgf/declare function={
		l(\m,\ns,\a,\mc)=1.604+\a*(1/\m-1/\mc)*\ns^(1/0.6301);
	},
	/pgf/declare function={
		lc(\m,\ns,\a,\mc,\b)=(1.604+\a*(1/\m-1/\mc)*\ns^(1/0.6301))*(1+\b*\ns^(-0.8940));
	},
	select coords between index/.style 2 args={
    	x filter/.code={
        	\ifnum\coordindex<#1\fi
        	\ifnum\coordindex>#2\fi
    	}
	}
}
\usepackage{pgfplotstable}
\usepackage{physics}

\newcommand{\clqcd}{\texttt{CL\kern-.25em\textsuperscript{2}QCD}}

\title{The QCD Deconfinement Critical Point for $N_\text{f}=2$ Flavors of Staggered Fermions}
\ShortTitle{The QCD Deconfinement Critical Point}

\author*[a,b]{Reinhold Kaiser}
\author[a,b]{Owe Philipsen}
\author[a,c]{Alessandro Sciarra}

\affiliation[a]{Institute for Theoretical Physics - Goethe University,\\
   Max-von-Laue-Str. 1,  60438 Frankfurt am Main, Germany}

\affiliation[b]{John von Neumann Institute for Computing (NIC) GSI,\\
Planckstr. 1, 64291 Darmstadt, Germany }

\affiliation[c]{Frankfurt Institute for Advanced Studies (FIAS) - Goethe University,\\
            Ruth-Moufang-Str. 1, 60438 Frankfurt am Main, Germany}

\emailAdd{kaiser@itp.uni-frankfurt.de}
\emailAdd{philipsen@itp.uni-frankfurt.de}
\emailAdd{sciarra@itp.uni-frankfurt.de}

\abstract{Quenched QCD at zero baryonic chemical potential undergoes a first-order
deconfinement phase transition at a critical temperature $T_c$, which is
related to the spontaneous breaking of the global center symmetry.
Including heavy, dynamical quarks breaks the center symmetry
explicitly and weakens the first-order phase transition. For
decreasing quark masses the first-order phase transition turns
into a smooth crossover at a $Z_2$-critical point. The critical quark
mass corresponding to this point has been examined with $N_\text{f} = 2$ Wilson
fermions for several $N_\tau$ in a recent study within our group. For
comparison, we also locate the critical point with $N_\text{f} = 2$
staggered fermions on $N_\tau = 8$ lattices. For this purpose we perform
Monte Carlo simulations for several quark mass values and various aspect
ratios in order to extrapolate to the thermodynamic limit. The critical
mass is obtained by fitting to a finite size scaling formula of the
kurtosis of the Polyakov loop. Our results indicate large discretization
effects, requiring simulations on lattices with $N_\tau > 8$.}

\FullConference{%
 The 38th International Symposium on Lattice Field Theory, LATTICE2021
  26th-30th July, 2021
  Zoom/Gather@Massachusetts Institute of Technology
}

\begin{document}
\maketitle

\section{Introduction}
\input{./sections/introduction}

\section{Simulation Details}
\input{./sections/simulation_details}

\section{Analysis of the Deconfinement Transition}
\input{./sections/analysis}

\section{Numerical Results}
\input{./sections/numerical_results}

\section{Conclusions}
\input{./sections/conclusions}

\acknowledgments{
The authors acknowledge support by the Deutsche Forschungsgemeinschaft (DFG, German Research Foundation) through the CRC-TR 211 \enquote{Strong-interaction matter under extreme conditions} – project number 315477589 – TRR 211 and by the State of Hesse within the Research Cluster ELEMENTS (Project ID 500/10.006).
We thank the Helmholtz Graduate School for Hadron and Ion Research (HGS-HIRe) for its support as well as the staff of L-CSC at GSI Helmholtzzentrum für Schwerionenforschung for computer time and support.
}

\bibliographystyle{JHEP}
\bibliography{main}

\end{document}

%% file: sections/introduction.tex
The nature of the thermal QCD transition at zero baryonic chemical potential has been found to be an analytic crossover for $N_\text{f}=2+1$ quark flavors with physical mass values \citep{aoki06}.
The sign problem prohibits the exploration of the QCD phase diagram at finite baryonic chemical potentials $\mu_B$ with lattice QCD.
However, investigating the nature of the thermal QCD transition at $\mu_B=0$ for varying quark mass values with lattice QCD gives valuable insights into the phase structure of QCD.
The nature of the thermal QCD transition at $\mu_B=0$ is visualized in the Columbia plot as a function of the degenerate up- and down-quark mass value $m_{u,d}$ and the strange quark mass value $m_s$.
First-order phase transition regions with a second-order boundary have been found with unimproved fermion discretizations on coarse lattices, for both heavy and light quark masses.
In contrast to the light quark mass regime, where strong indications exist that the first order region vanishes towards the continuum limit \citep{cuteri21}, the first-order region in the heavy mass regime is known to persist from investigations of pure gauge theory \citep{boyd96}.
A recent study with $N_\text{f}=2$ unimproved Wilson fermions even shows an enlargement of the first-order region when approaching the continuum limit \citep{cuteri20}. 

The deconfinement transition in the heavy quark mass regime, which is subject of this work, is related to the spontaneous breaking of the $Z_3$ center symmetry.
In the limit of infinite quark mass values, QCD is invariant under $Z_3$ center symmetry transformations, leading to a first-order phase transition at the transition temperature $T_c$.
The inclusion of large but finite quark masses breaks the center symmetry explicitly. 
Decreasing the quark masses weakens the first-order phase transition until the deconfinement transition turns into an analytic crossover at a $Z_2$ second-order boundary.
This $Z_2$-critical point has been investigated for $N_\text{f}=2$ quark flavors and 3 different temporal lattice extents $N_\tau$ employing the Wilson fermion action \citep{cuteri20}.
For comparison, the goal is to locate the same $Z_2$-critical point employing the unimproved staggered fermion action.
As a first step, we present results for $N_\tau=8$ in this work.

The motivation to study the thermal QCD transition far from the physical quark mass values is to provide a first-principles benchmark for effective theories, that are not limited by the sign problem and can access the $\mu_B\neq 0$ region.
These effective theories include effective lattice theories obtained from the hopping parameter expansion \citep{fromm12, saito14, aarts16} and effective Polyakov loop theories in the continuum \citep{fischer15, lo14}.
Furthermore, the $Z_2$-critical point is of interest, as it is the point where the latent heat of the first-order deconfinement phase transition vanishes.	
Hence, the dynamics of the deconfinement transition relates the value of the critical quark mass and the latent heat such that non-perturbative investigations allow valuable insights in this parameter region.
For pure gauge theory with infinite quark mass values, the latent heat has already been calculated \citep{shirogane16}.

%% file: sections/simulation_details.tex
The Monte Carlo importance sampling simulations are preformed using the standard Wilson gauge action
\begin{equation}\label{equ:wilson-gauge-action}
S_g = \frac{\beta}{3}\sum_n\sum_{\mu<\nu}\Re\left(\Tr\left[\mathbb{1}-P_{\mu\nu}(n)\right]\right),
\end{equation}
with the inverse gauge coupling $\beta=\frac{6}{g^2}$, the plaquette $P_{\mu\nu}(n)$ depending on the lattice sites $n$ and directions $\mu$ and $\nu$.
For the fermions, the staggered fermion action
\begin{equation}
S_f = \sum_n \bar{\psi}(n)\left(\sum_{\mu=1}^4\eta_\mu(n)\frac{U_\mu(n)\psi(n+\hat\mu)-U^\dagger_\mu(n-\hat\mu)\psi(n-\hat\mu)}{2}+(am)\psi(n)\right)
\end{equation}
is employed, where $a$ is the lattice spacing and $\bar\psi$ and $\psi$ are the fermion fields with mass $am$.
The gauge links are given by $U_\mu(n)$ and $\eta_\mu(n)$ is the staggered sign function.
The inverse gauge coupling $\beta$ controls the lattice spacing $a(\beta)$ and the temperature $T=\frac{1}{a(\beta)N_\tau}$, where $N_\tau=8$ is the number of lattice points in temporal direction. 
The number of lattice points in spatial direction $N_\sigma$ defines the size of the system $L=a(\beta)N_\sigma$, for which five aspect ratios $LT=N_\sigma/N_\tau\in[4,8]$ were simulated.
Simulations were run for two or three values of $\beta$ around the (pseudo-)critical $\beta_c$ at which the transition takes place.
After a thermalization process, four independent Markov chains are produced for each $\beta$ value.
In order to locate the critical quark mass $m_c$ at the $Z_2$-critical point, simulations were run for the mass values $am\in\{0.35,0.55,0.75,0.95,1.15\}$. 

The order parameter associated with the deconfinement transition is the expectation value of the Polyakov loop $L$, averaged over the spatial lattice volume, with
\begin{equation}
L=\frac{1}{N_\sigma^3}\sum_{\bm n}\frac{1}{3}\Tr\left[\prod_{n_0=0}^{N_\tau-1}U_0(n_0, \bm n)\right].
\end{equation}
The product is performed over all temporal gauge links at spatial lattice site $\bm n$, closing through the periodic boundary in temporal direction.

After $5000$ trajectories of thermalization, an amount of 200k to 300k trajectories was accumulated for each Markov chain, which sums up to a maximum of 1.2M trajectories for one $\beta$ value.
The configurations are generated by the RHMC algorithm, where the acceptance rate is kept at $\sim 85\%$.
The employed lattice QCD code is the Open-CL based code \clqcd\ \citep{bach13}, also publicly available \citep{pinke18}, which is run on GPUs on the L-CSC cluster at GSI in Darmstadt, Germany.
To manage the simulations the bash tool \texttt{BaHaMAS} was used, which allows to submit and monitor huge amounts of simulations efficiently \citep{sciarra21}.

%% file: sections/analysis.tex
We use the standardized moments skewness, $B_3$, and kurtosis, $B_4$, of the norm of the Polyakov loop to localize the deconfinement transition and determine its order.
The $n$-th standardized moment of the expectation value of the absolute value of the Polyakov loop $\abs{L}$ with mean value $\overline{\abs{L}}$ is given by
\begin{equation}
B_n = \frac{\expval{\left(\abs{L}- \overline{\abs{L}}\right)^n}}{\expval{\left(\abs{L} - \overline{\abs{L}}\right)^2}^{\frac{n}{2}}}.
\end{equation}
The skewness measures the asymmetry of the distribution and serves to locate the (pseudo-)critical $\beta_c$ using the condition $B_3(\beta_c)=0$.
The kurtosis of the distribution at $\beta_c$ gives information about the type of the transition, which assume specific values in the infinite volume limit (see Table \ref{tab:kurtosis-values}).

\begin{table}
\centering
\begin{tabular}{|l|c|c|c|}
\hline
\textbf{Transition type} & 1. Order & $Z_2$ (Ising 3D) \citep{blote95} & Crossover\\
\hline
$\bm{B_4(\beta_c)}$ & $1$ & $1.604(1)$& $3$\\
\hline
\end{tabular}
\caption{Infinite volume kurtosis values of the order parameter for selected transitions.\label{tab:kurtosis-values}}
\end{table}

\begin{figure}
\input{tikz/analysis_reweighting_plot}
\caption{Exemplary analysis and reweighting of $B_3$ and $B_4$ of $\abs{L}$ for $am=0.55$ and $N_\sigma=56$.
The points on the left are shifted horizontally around the central value for readability. 
The upper row shows the skewness, the lower row shows the kurtosis.
The left column shows the analysis of the independent Markov chains. 
The analysis of the merged chains is shown on the right in red, as well as the reweighted data points.\label{fig:analysis}}
\end{figure}
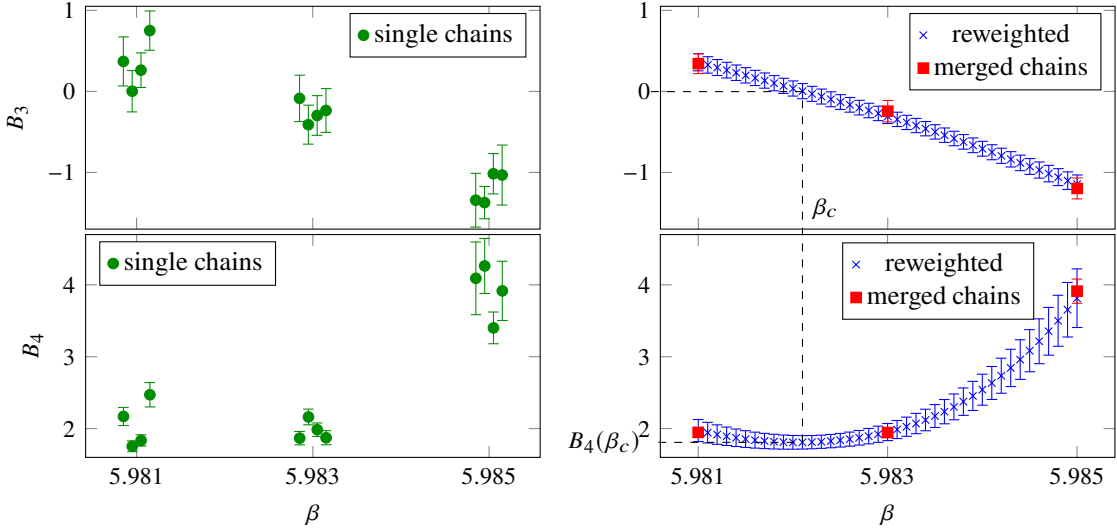

The sampling of the simulated $\beta$ values is too coarse, such that $\beta_c$ cannot be extracted directly.
Interpolating the skewness and the kurtosis in between the simulated $\beta$ values using the multiple histogram method \citep{ferrenberg89} leads to good estimates of $\beta_c$ and $B_4(\beta_c)$.
This analysis strategy is visualized in Figure \ref{fig:analysis} for the parameters $am=0.55$ and $N_\sigma = 56$.

The set of kurtosis values at $\beta_c$ obtained from this analysis is volume dependent and a finite size scaling analysis has to be performed.
The finite size scaling formula for the kurtosis of $\abs{L}$, which is in general a mixture of an energy- and magnetic-like observable, is
\begin{equation}\label{equ:kurtosis-finite-size}
B_4(N_\sigma, \beta_c, am)=\left(A+B x + 
\order{x^2}\right)\cdot \left( 1 + C N_\sigma^{y_t-y_h} + \order{N_\sigma^{2(y_t{-}y_h)}}\right),
\end{equation}
with the scaling variable $x=\left(\frac{1}{am}-\frac{1}{am_c}\right)N_\sigma^{1/\nu}$ \citep{takeda16}.
The critical exponents $y_t=1/\nu=1.5870(10)$ and $y_h=2.4818(3)$ are known for the $Z_2$ universality class \citep{pelissetto02}.
$A, B, C$ are undetermined constants from the Taylor expansions and $am_c$ is the value of the $Z_2$-critical mass in lattice units.
Based on formula \eqref{equ:kurtosis-finite-size}, two fit ansätze can be derived.
The general one includes a correction term
\begin{equation}\label{equ:fit-with-correction}
B_4(N_\sigma, am) = \left(1.604 + b_1 \left(\frac{1}{am}-\frac{1}{am_c}\right)N_\sigma^{1/\nu}\right)\cdot \left( 1+ cN_\sigma^{y_t-y_h}\right),
\end{equation}
whereas for sufficiently large volumes $N_\sigma$ the correction term can be neglected,
\begin{equation}\label{equ:fit-without-correction}
B_4(N_\sigma, am) = 1.604 + b_1 \left(\frac{1}{am}-\frac{1}{am_c}\right)N_\sigma^{1/\nu}.
\end{equation}
By fitting to these fit formulas, the critical mass $am_c$ can be extracted as a fit parameter besides the other fit parameters $b_1$ and $c$.

%% file: tikz/analysis_reweighting_plot.tex
\begin{tikzpicture}
\def\a{6.5}

\begin{axis}[ylabel={$B_3$},xticklabels={}, x tick label style={/pgf/number format/.cd, fixed, precision=3}, ymin=-1.7, ymax=1.05, legend style = {font=\small, anchor = north east, legend pos = north east}, height=0.3\textwidth, width=0.5\textwidth, xtick={5.981, 5.983, 5.985}]
\addplot+[error bars/.cd, y dir=both, y explicit, /tikz/.cd, only marks,mark size=0.7mm,green!55!black, mark options={fill=green!55!black},] table[x index=0, y index=1, y error index=2]{./anc/multiple_chains_skewness.dat};
\addlegendentry{single chains};
\end{axis}

\begin{scope}[xshift=\textwidth/2]
\begin{axis}[xticklabels={}, x tick label style={/pgf/number format/.cd, fixed, precision=3}, ymin=-1.7, ymax=1.05, legend style = {font=\small, anchor = north east, legend pos = north east}, height=0.3\textwidth, width=0.5\textwidth,xtick={5.981, 5.983, 5.985}]
\addplot+[error bars/.cd, y dir=both, y explicit, /tikz/.cd, only marks,mark=x,mark size=0.7mm,blue] table[x index=0, y index=1, y error index=2]{./anc/reweighted_skewness.dat};
\addlegendentry{reweighted};
\addplot+[error bars/.cd, y dir=both, y explicit, /tikz/.cd, only marks, mark size=0.7mm,red, mark options={fill=red},] table[x index=0, y index=1, y error index=2]{./anc/merged_chains_skewness.dat};
\addlegendentry{merged chains};
\node (zero) at (axis cs:5.9805,0) {};
\node (betaca) at (axis cs:5.9821,0) {};
\node[anchor= south west] at (axis cs:5.9821,-1.7) {$\beta_c$};
\end{axis}
\end{scope}

\begin{scope}[yshift=-0.20\textwidth]
\begin{axis}[xlabel={$\beta$}, ylabel={$B_4$},x tick label style={/pgf/number format/.cd, fixed, precision=3}, ymin=1.6, ymax=4.7, legend style = {font=\small, anchor = north west, legend pos = north west}, height=0.3\textwidth, width=0.5\textwidth,xtick={5.981, 5.983, 5.985}]
\addplot+[error bars/.cd, y dir=both, y explicit, /tikz/.cd, only marks,mark size=0.7mm,green!55!black, mark options={fill=green!55!black},] table[x index=0, y index=1, y error index=2]{./anc/multiple_chains_kurtosis.dat};
\addlegendentry{single chains};
\end{axis}
\end{scope}

\begin{scope}[yshift=-0.20\textwidth, xshift=\textwidth/2]
\begin{axis}[xlabel={$\beta$}, x tick label style={/pgf/number format/.cd, fixed, precision=3}, ymin=1.6, ymax=4.7, legend style = {font=\small, anchor = north west, at={(0.4,0.965)}}, height=0.3\textwidth, width=0.5\textwidth,xtick={5.981, 5.983, 5.985}]
\addplot+[error bars/.cd, y dir=both, y explicit, /tikz/.cd, only marks, mark=x,mark size=0.7mm, blue] table[x index=0, y index=1, y error index=2]{./anc/reweighted_kurtosis.dat};
\addlegendentry{reweighted};
\addplot+[error bars/.cd, y dir=both, y explicit, /tikz/.cd, only marks, mark size=0.7mm,red, mark options={fill=red},] table[x index=0, y index=1, y error index=2]{./anc/merged_chains_kurtosis.dat};
\addlegendentry{merged chains};
\node (betacb) at (axis cs:5.9821,1.809082078463) {};
\node (bfour) at (axis cs:5.9805,1.809082078463) {};
\end{axis}
\end{scope}
\draw[dashed] (zero.center) -- (betaca.center) -- (betacb.center) -- (bfour.center);
\node[anchor = east] at (bfour) {$B_4(\beta_c)$};
               
\end{tikzpicture}

%% file: sections/numerical_results.tex
\begin{figure}
\input{tikz/fitting-plots}
\caption{Kurtosis fits without and with correction term.
The kurtosis points are shifted horizontally around the central value for readability.
(a) shows a linear fit without correction (see formula \eqref{equ:fit-without-correction}), where all points from $N_\sigma=32$ and one point from $N_\sigma=40$ are excluded.
(b) shows a linear fit with correction (see formula \eqref{equ:fit-with-correction}), where all points are included. 
The value of the critical quark mass is indicated by the black line and its error by the gray area.
\label{fig:kurtosis-fits}
}
\end{figure}
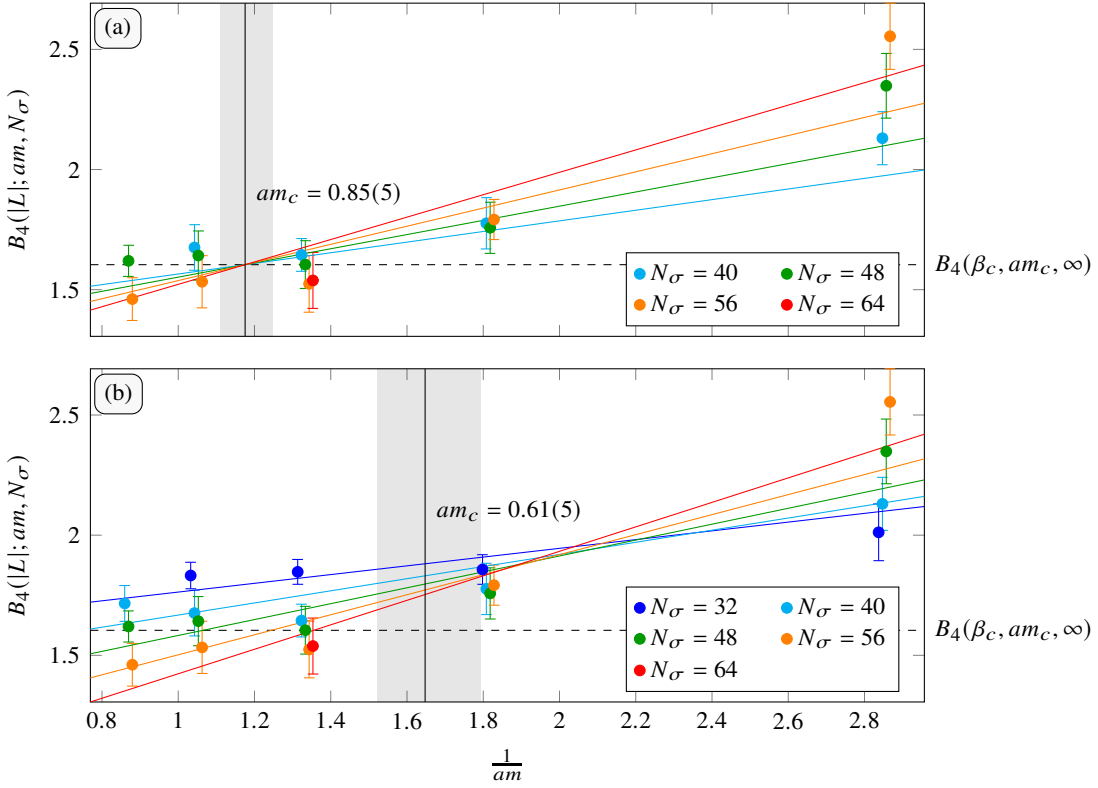

Having obtained one kurtosis value at $\beta_c$ for each simulated combination of $N_\sigma$ and $am$ as described in the previous section, both fit formulas \eqref{equ:fit-with-correction} and \eqref{equ:fit-without-correction} are fitted to this set of data.
A good fit without correction is obtained by excluding the kurtosis points for the smallest volume $N_\sigma=32$ and one point from $N_\sigma=40$.
The result is shown in Figure \ref{fig:kurtosis-fits} (a).
By applying the fit with correction, all simulated kurtosis points can be included and a better fit is obtained.
The corresponding plot is shown in Figure \ref{fig:kurtosis-fits} (b).
The numerical fit results for both fits are shown in Table \ref{tab:fit-results}.

\begin{table}
\centering
\begin{tabular}{|l|l|l|l|l|l|l|}
\hline
& $m_c$ & $b_1$ & $c$ & ndf & $\chi^2_\text{ndf}$ & $Q$\\\hline
fit without correction (a) & $0.85(5)$ & $0.00079(9)$ & - & $13$ & $1.13$ & $32.4\%$\\\hline
fit with correction (b) & $0.61(5)$ & $0.00063(7)$ & $3.8(6)$ & $17$ & $1.03$ & $42.1\%$ \\\hline
\end{tabular}
\caption{Fit results for the two fits displayed in Figure \ref{fig:kurtosis-fits}.
$m_c, b_1, c$ are fit parameters given with their errors in parentheses.
$\text{ndf}$ is the number of degrees of freedom and $\chi^2_\text{ndf}$ is the $\chi^2$ per number of degrees of freedom.
The quality of the fit is quantified by the $Q$-parameter.
\label{tab:fit-results}}
\end{table}

Comparing the plots of both fits in Figure \ref{fig:kurtosis-fits}, the correction term shifts the pairwise crossing point of the kurtosis lines up.
This leads to a better fit, which is expressed in the better $\chi^2_\text{ndf}$ and fit quality parameter $Q$ in Table \ref{tab:fit-results}.
Especially the kurtosis points for the smallest mass value are hit much better by the fit function with correction.
The results for the critical mass $am_c$ are not compatible with each other, indicating a still large finite size effect, even if the smallest volume is excluded from the fit.
This statement is supported by the result of the fits with correction term, which consecutively exclude the smaller volumes.
The dependence on the correction term parameter $c$ persists, as it does not become compatible with zero when only including the largest volumes.

\begin{table}
\centering
\begin{tabular}{|l|l|l|l|l|l|}
\hline
$am$ & $\beta_c$ & $am_\pi$ & $a\; \{\unit{fm}\}$ & $m_\pi\; \{\unit{GeV}\}$ & $T_c\; \{MeV\}$ \\
\hline
$0.55$ & $5.9821$ & $1.72039(7)$ & $0.0888(10)$ & $3.82(4)$ & $278(3)$ \\
$0.75$ & $6.0129$ & $1.98121(7)$ & $0.0872(9)$ & $4.48(5)$ & $283(3)$ \\
\hline
\end{tabular}
\caption{Results from pion mass measurement and scale setting. \label{tab:pion-scale-results}}
\end{table}

Regarding the considerations from above, the fit with correction term is chosen to be the best fit and the final result for the critical mass of the $Z_2$ critical point on $N_\tau=8$ lattices is
\begin{equation}\label{equ:final-critical-mass}
am_c=0.61(5).
\end{equation}
The pion masses for the simulated quark masses are calculated at $\beta_c$ on $N_\tau=32$ and $N_s=16$ lattices.
Furthermore, the scale is set using the $w_0$-scale \citep{borsanyi12} based on the Wilson flow \citep{luescher10}.
The results are shown in Table \ref{tab:pion-scale-results} for the two neighboring simulated masses around the critical mass.
The pion mass in lattice units, which is larger than $1$, contains significant cutoff effects, as the lattice is not able to resolve the pion represented by its Compton wavelength $1/m_\pi$.

\begin{wraptable}[9]{r}{5.2cm}
\centering
\begin{tabular}{|l|r|}
\hline
$N_\tau$ & $m_\pi^{Z_2}$ $\{\unit{GeV}\}$\\
\hline
$6$ & $5.01(5)$\\
$8$ & $4.51(5)$\\
$10$ & $4.39(5)$\\
\hline
\end{tabular}
\caption{Values of the critical pseudo scalar meson mass $m_\pi^{Z_2}$ for various $N_\tau$ from Wilson fermions \citep{cuteri20}.\label{tab:mpi_wilson}}
\end{wraptable}

For comparison, the results from Wilson fermions for three different $N_\tau=6,8,10$ are shown in Table \ref{tab:mpi_wilson}.
The determination of the critical pion mass for the critical quark mass value in equation \eqref{equ:final-critical-mass} for staggered fermions on $N_\tau=8$ lattices has not yet been performed, as it is to early to use this value for a quantitative continuum extrapolation.
However, the measured pion masses in Table \ref{tab:pion-scale-results} are roughly consistent with the results from Wilson fermions on $N_\tau=8$ lattices.

%% file: tikz/fitting-plots.tex
\begin{tikzpicture}
\pgfplotstableread{anc/kurtosis_points.dat}{\kurtosistable};
\begin{axis}[width=\textwidth*0.83, height=6cm, enlargelimits=false, legend pos=south east, legend columns=2, legend style={/tikz/every even column/.append style={column sep=0.5cm}}, xticklabels={}, ylabel={$B_4(\abs{L};am, N_\sigma)$}, /pgfplots/ymin=1.3058195, clip=false]
\path[fill=black,draw=none, opacity=0.1]
(axis cs:1/0.9009,\pgfkeysvalueof{/pgfplots/ymin}) -- 
(axis cs:1/0.9009,\pgfkeysvalueof{/pgfplots/ymax}) -- 
(axis cs:1/0.8007,\pgfkeysvalueof{/pgfplots/ymax}) -- 
(axis cs:1/0.8007,\pgfkeysvalueof{/pgfplots/ymin}) -- cycle;
\addplot [black, dashed, domain=1/1.15-0.1:1/0.35+0.1, forget plot]{1.604} node [pos=1, anchor=west] {$B_4(\beta_c, am_c, \infty)$};
\addplot [
	cyan,
	select coords between index={4}{7},
	only marks,
	error bars/.cd, y dir = both, y explicit,
] table [x expr=1/\thisrowno{0}-0.01, y=Kurtosis, y error=errorKurtosis] {\kurtosistable};
\addplot [
	green!60!black,
	select coords between index={9}{13},
	only marks,
	error bars/.cd, y dir = both, y explicit,
] table [x expr=1/\thisrowno{0}, y=Kurtosis, y error=errorKurtosis] {\kurtosistable};
\addplot [
	orange,
	select coords between index={14}{18},
	only marks,
	error bars/.cd, y dir = both, y explicit,
] table [x expr=1/\thisrowno{0}+0.01, y=Kurtosis, y error=errorKurtosis] {\kurtosistable};
\addplot [
	red,
	select coords between index={19}{19},
	only marks,
	error bars/.cd, y dir = both, y explicit,
] table [x expr=1/\thisrowno{0}+0.02, y=Kurtosis, y error=errorKurtosis] {\kurtosistable};
\legend{$N_\sigma=40$, $N_\sigma=48$, $N_\sigma=56$, $N_\sigma=64$};
\addplot [cyan, domain=1/1.15-0.1:1/0.35+0.1]{l(1/x, 40, 0.0006341, 0.8508)};
\addplot [green!60!black, domain=1/1.15-0.1:1/0.35+0.1]{l(1/x, 48, 0.0006341, 0.8508)};
\addplot [orange, domain=1/1.15-0.1:1/0.35+0.1]{l(1/x, 56, 0.0006341, 0.8508)};
\addplot [red, domain=1/1.15-0.1:1/0.35+0.1]{l(1/x, 64, 0.0006341, 0.8508)};
\draw (axis cs:1/0.8508,\pgfkeysvalueof{/pgfplots/ymin}) -- (axis cs:1/0.8508,\pgfkeysvalueof{/pgfplots/ymax});
\node[anchor = west] at (axis cs: 1.178, 1.9) {$am_c=0.85(5)$};
\node[rectangle, draw, anchor=north west,fill=black!3!white, rounded corners] at (axis cs:\pgfkeysvalueof{/pgfplots/xmin}+0.01,\pgfkeysvalueof{/pgfplots/ymax}-0.02) {(a)};
\end{axis}
\end{tikzpicture}

\begin{tikzpicture}
\pgfplotstableread{anc/kurtosis_points.dat}{\kurtosistable};
\begin{axis}[width=\textwidth*0.83, height=6cm, enlargelimits=false, legend pos=south east, legend columns=2, legend style={/tikz/every even column/.append style={column sep=0.5cm}}, xlabel=$\frac{1}{am}$, ylabel={$B_4(\abs{L}; am, N_\sigma)$}, clip=false]
\path[fill=black,draw=none, opacity=0.1]
(axis cs:1/0.5572,\pgfkeysvalueof{/pgfplots/ymin}) --
(axis cs:1/0.5572,\pgfkeysvalueof{/pgfplots/ymax}) -- 
(axis cs:1/0.6568,\pgfkeysvalueof{/pgfplots/ymax}) --
(axis cs:1/0.6568,\pgfkeysvalueof{/pgfplots/ymin}) -- cycle;
\addplot [black, dashed, domain=1/1.15-0.1:1/0.35+0.1, forget plot]{1.604} node [pos=1, anchor=west] {$B_4(\beta_c, am_c, \infty)$};
\addplot [
	blue,
	select coords between index={0}{3},
	only marks,
	error bars/.cd, y dir = both, y explicit,
] table [x expr=1/\thisrowno{0}-0.02, y=Kurtosis, y error=errorKurtosis] {\kurtosistable};
\addplot [
	cyan,
	select coords between index={4}{8},
	only marks,
	error bars/.cd, y dir = both, y explicit,
] table [x expr=1/\thisrowno{0}-0.01, y=Kurtosis, y error=errorKurtosis] {\kurtosistable};
\addplot [
	green!60!black,
	select coords between index={9}{13},
	only marks,
	error bars/.cd, y dir = both, y explicit,
] table [x expr=1/\thisrowno{0}, y=Kurtosis, y error=errorKurtosis] {\kurtosistable};
\addplot [
	orange,
	select coords between index={14}{18},
	only marks,
	error bars/.cd, y dir = both, y explicit,
] table [x expr=1/\thisrowno{0}+0.01, y=Kurtosis, y error=errorKurtosis] {\kurtosistable};
\addplot [
	red,
	select coords between index={19}{19},
	only marks,
	error bars/.cd, y dir = both, y explicit,
] table [x expr=1/\thisrowno{0}+0.02, y=Kurtosis, y error=errorKurtosis] {\kurtosistable};
\legend{$N_\sigma=32$, $N_\sigma=40$, $N_\sigma=48$, $N_\sigma=56$, $N_\sigma=64$};
\addplot [blue, domain=1/1.15-0.1:1/0.35+0.1]{lc(1/x, 32, 0.0006341, 0.607, 3.827)};
\addplot [cyan, domain=1/1.15-0.1:1/0.35+0.1]{lc(1/x, 40, 0.0006341, 0.607, 3.827)};
\addplot [green!60!black, domain=1/1.15-0.1:1/0.35+0.1]{lc(1/x, 48, 0.0006341, 0.607, 3.827)};
\addplot [orange, domain=1/1.15-0.1:1/0.35+0.1]{lc(1/x, 56, 0.0006341, 0.607, 3.827)};
\addplot [red, domain=1/1.15-0.1:1/0.35+0.1]{lc(1/x, 64, 0.0006341, 0.607, 3.827)};

\draw (axis cs:1/0.607,\pgfkeysvalueof{/pgfplots/ymin}) -- (axis cs:1/0.607,\pgfkeysvalueof{/pgfplots/ymax});
\node[anchor = west] at (axis cs: 1.65, 2.1) {$am_c=0.61(5)$};
\node[rectangle, draw, anchor=north west,fill=black!3!white, rounded corners] at (axis cs:\pgfkeysvalueof{/pgfplots/xmin}+0.01,\pgfkeysvalueof{/pgfplots/ymax}-0.02) {(b)};
\end{axis}
\end{tikzpicture}

%% file: sections/conclusions.tex
This work presents a first step to determine the heavy $Z_2$-critical quark mass at zero chemical potential for $N_\text{f}=2$ flavors with staggered fermions.
The analysis of the fits with and without the correction term has shown a significant dependence on the correction term for obtaining good fits.
To avoid the use of the correction term, simulations on much larger spatial lattices would be needed.
Currently, this is not feasible as the computing time per trajectory would grow as well as the autocorrelation time of the observables due to the critical slowing down.
The fact that only one $N_\tau$ has been fully simulated up to this point does not allow any conclusions about the expected shift of the critical quark mass towards smaller values for increasing $N_\tau$.
The large values of the pion masses in lattice units also imply large discretization effects and the necessity to simulate on finer lattices.